\documentclass[preprint,12pt]{elsarticle}
\makeatletter
\def\ps@pprintTitle{%
  \let\@oddhead\@empty
  \let\@evenhead\@empty
  \let\@oddfoot\@empty
  \let\@evenfoot\@oddfoot
}
\makeatother

\usepackage{amssymb}
\usepackage{amsmath}
\usepackage{color}
\usepackage[normalem]{ulem}
\usepackage{hyperref}
\hypersetup{colorlinks=true,citecolor=red,linkcolor=blue}

\begin{document}

\begin{frontmatter}



\title{Spatio-temporal Pattern Formation due to Host-Circuit Interplay in Gene Expression Dynamics}


\author[inst1]{Priya Chakraborty}

\affiliation[inst1]{organization={Department of Physics},
            addressline={National Institute of Technology}, 
            city={Durgapur},
            postcode={713209}, 
            state={West Bengal},
            country={India}}

\author[inst2]{Mohit K. Jolly}
\author[inst2]{Ushasi Roy\footnote{Corresponsing author: \textit{ushasiroy@iisc.ac.in}}}
\author[inst1]{Sayantari Ghosh\footnote{Corresponsing author: \textit{sayantari.ghosh@phy.nitdgp.ac.in} }}

\affiliation[inst2]{organization={ Centre for BioSystems Science and Engineering},
            addressline={Indian Institute of Science}, 
            city={Bangalore},
            postcode={560012}, 
            state={Karnataka},
            country={India}}
\begin{abstract}
Biological systems are majorly dependent on their property of bistability in order to exhibit nongenetic heterogeneity in terms of cellular morphology and physiology. Spatial patterns of phenotypically heterogeneous cells, arising due to underlying bistability, may play significant role in phenomena like biofilm development, adaptation, cell motility etc. While nonlinear positive feedback regulation, like cooperative heterodimer formation are the usual reason behind bistability,  similar dynamics can also occur as a consequence of host-circuit interaction. In this paper, we have investigated the pattern formation by a motif with non-cooperative positive feedback, that imposes a metabolic burden on its host due to its expression. In a cellular array set inside diffusible environment, we investigate spatio-temporal diffusion in one dimension as well as in two dimension in the context of various initial conditions respectively. Moreover, the number of cells exhibiting the same steady state, as well as their spatial distribution has been quantified in terms of connected component analysis. The effect of diffusion coefficient variation has been studied in terms of stability of related states and time evolution of patterns.
\end{abstract}




\end{frontmatter}


\section{Introduction}
\label{intro}
Proteins 
are responsible for 
diverse functionalities, serving the cells for structural support, motility, enzymatic activity, inner organization, interaction with the outside environment and many more \cite{orengo1999protein}. From DNA to mRNA, and then to proteins, the information flows in a tightly controlled way inside cell. This flow of information, i.e., gene expression 
has two major steps: \textit{transcription} and \textit{translation}, taking places in cell nucleus and cytoplasm respectively. 
Transcriptional gene regulation is one of the fundamental ways that control expression of any particular gene in terms of location, amount and timing. Though all the cells in a isogenic microbial population contains same genome, switching the expression of a gene ON and OFF, the cell population can get bifurcated into two subpopulations, which are distinct, but coexisting. Bistability, a vastly preferred physical behavior of living cells, is known to regulate this ``nongenetic", phenotypic heterogeneity \cite{veening2008bistability, dubnau2006bistability,grote2015phenotypic}.
In bistable response, protein concentration attains any of the two drastically different steady states (low or high response), and shows a history-dependent behaviour, \textit{hysteresis}. The phenomena introduces a memory in the system, making it retain its  state instead of variation or fluctuations in inducer level. 
\\The essential nonlinearity required for bistability is conventionally achieved by a genetic system through a positive feedback with cooperative regulation by multimer formation, or by a combination of more than one feedback loops. 
However, recently, bistability driven by host-circuit coupling have been depicted by researchers \cite{tan2009emergent,melendez2022emergence}.  In \cite{tan2009emergent}, the authors termed this phenomena as \textit{emergent bistability}, where, along with a positive non-cooperative gene regulation, a secondary double-negative feedback loop provides the necessary nonlinearity, indirectly originating due to host-circuit coupling. While toxicity of expressed protein may  be a reason behind growth retardation of host cell \cite{klumpp2009growth}, this phenomena can be much prevalent in natural systems also \cite{lewis2007persister,ghosh2011phenotypic, young2008systems}. A possible explanation lies in the fact that in  presence of limited resources, the  protein synthesis may impose a metabolic burden, causing a reduction in the amount of resources available for cellular growth. 
\\
The connection between spatially and/or temporally structured phenotypic  heterogeneity with  diversification and adaptation of populations have been explored in several works \cite{agrawal2001phenotypic,allen2012trait,michel2011spatial}. Though, in most of the related works, environmental fluctuations and variations are associated with this patterning, in his seminal work, A. Turing had already shown that two interacting diffusing chemicals can generate a stable inhomogeneous pattern, under certain conditions \cite{turing1990chemical}. Connecting this biological systems, mathematical models have been formed to explore  the dynamics of pattern formation in Activator-Inhibitor system \cite{gierer1972theory, othmer1971instability, lengyel1992chemical, cross1993pattern}, feedback quenched oscillator system \cite{hsia2012feedback} and many more \cite{miyazako2013turing}. Eventually successful experimental demonstrations have provided a strong background to these theoretical models \cite{zaikin1970concentration, winfree1972spiral, castets1990experimental}. However, the lesser explored area of pattern formation in gene regulatory systems has drawn the attention of research community very recently and different mechanisms are employed to study realistic scenarios \cite{diambra2015cooperativity, perez2018combining,  barbier2020controlling, luo2019synthetic,roy2022spatiotemporal}. 
\\ In this work, we explore the effect of the inherent stochasticity due to spatial diffusion, in presence of host-circuit coupling to observe spatially and temporally structured pattern formation. For this purpose we consider the emergent bistable dynamics we have mentioned before \cite{tan2009emergent}.  Though, for this dynamics, successful implementation with synthetic gene circuits have been achieved and stochastic responses have also been studied \cite{tan2009emergent, ghosh2012emergent}, no significant study on the pattern formation in a diffusive environment by the system is done as per our knowledge. Moreover, in all these studies a single, isolated genetic circuit has been considered along with its host cell. On the other hand, a collection of host cells, each with the circuit of interest embedded, evolving in presence of diffusion of the synthesized protein, is a scenario more closely relatable with experiments. So, we investigate the spatio-temporal behaviour in this thorough study. Here, in this paper we described the model formulation in Sec. \ref{model}, deterministic analyses related to equilibria and bifurcation is described in Sec. \ref{res:bifur} and the detailed results of reaction diffusion system is explained in Sec. \ref{res diff}. Further quantitative analysis of this host-circuit interaction driven spatio-temporal pattern formation is done in Sec. \ref{quanti}. Finally, in Sec. \ref{diss}, we concluded with brief discussion on relevance of the present work and future perspective. 
\section{Model formulation}\label{model}
\begin{figure}
    \centering
    \includegraphics[width=0.45\textwidth]{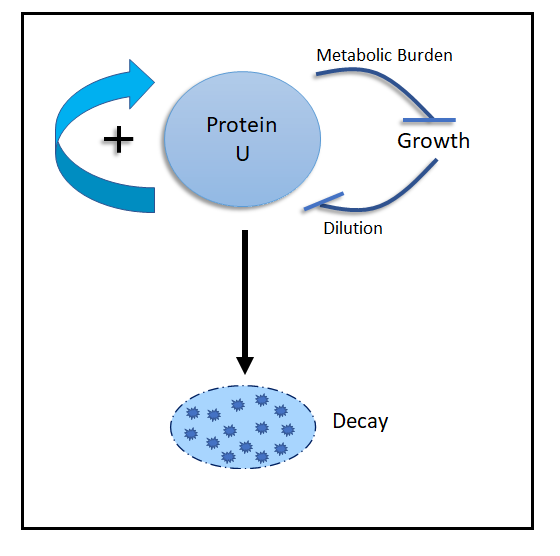}
    \caption{Schematic diagram of model motif which shows emergent bistability. Protein $U$ activates its own synthesis by a positive feedback loop (thick blue arrow) and has a natural decay rate (black arrow). Synthesis of the protein puts a metabolic burden affecting cellular growth.  As growth is hindered, protein dilution gets reduced creating a double negative feedback loop (marked by consecutive hammerheads). This acts effectively as a positive feedback.}
    \label{motif}
\end{figure}
Let us describe the model formulation of the concerned motif as shown in Fig. \ref{motif}. Let $U$ be a protein which activates its own synthesis, with an effective synthesis rate constant, $\alpha$. We consider no coopertivity associated in this positive feedback. Basal synthesis rate of the protein $U$ is represented by $\delta$. Growth causing the increase in volume dilutes the protein, we denote this dilution rate can achieve the maximum value $\phi$. Now, to incorporate host-circuit interaction, we consider that the expression of protein $U$ is associated with an expense of resources present in cell. This effectively creates a metabolic burden on cell growth affecting the protein dilution. We consider $\beta$ and $\gamma$ to be the linear and nonlinear reduction in dilution rate due to this host-circuit coupling.  The natural degradation rate has been taken into account by $\Delta_U$. The idea of considering nonlinear degradation as a consequence of metabolic burden has been explored in other works  \cite{tan2009emergent,melendez2022emergence}  and briefly explained by Monod in \cite{monod2012growth}. Another valid explanation includes the possibilities of direct/indirect toxic effect of protein synthesis on growth of the cell \cite{klumpp2009growth}. Now, the form of mathematical representation of the above considerations are given by Eq. \ref{eqn main}.
\begin{equation}\label{eqn main}
    \frac{dU}{dt}= \frac{\delta + \alpha\:U}{1+U} -\frac{\phi\:U}{\beta+\gamma\:U}-\Delta_U\:U
\end{equation}
We proceed with further analysis with this model equation.
\begin{figure}
    \centering
    \includegraphics[width=\textwidth]{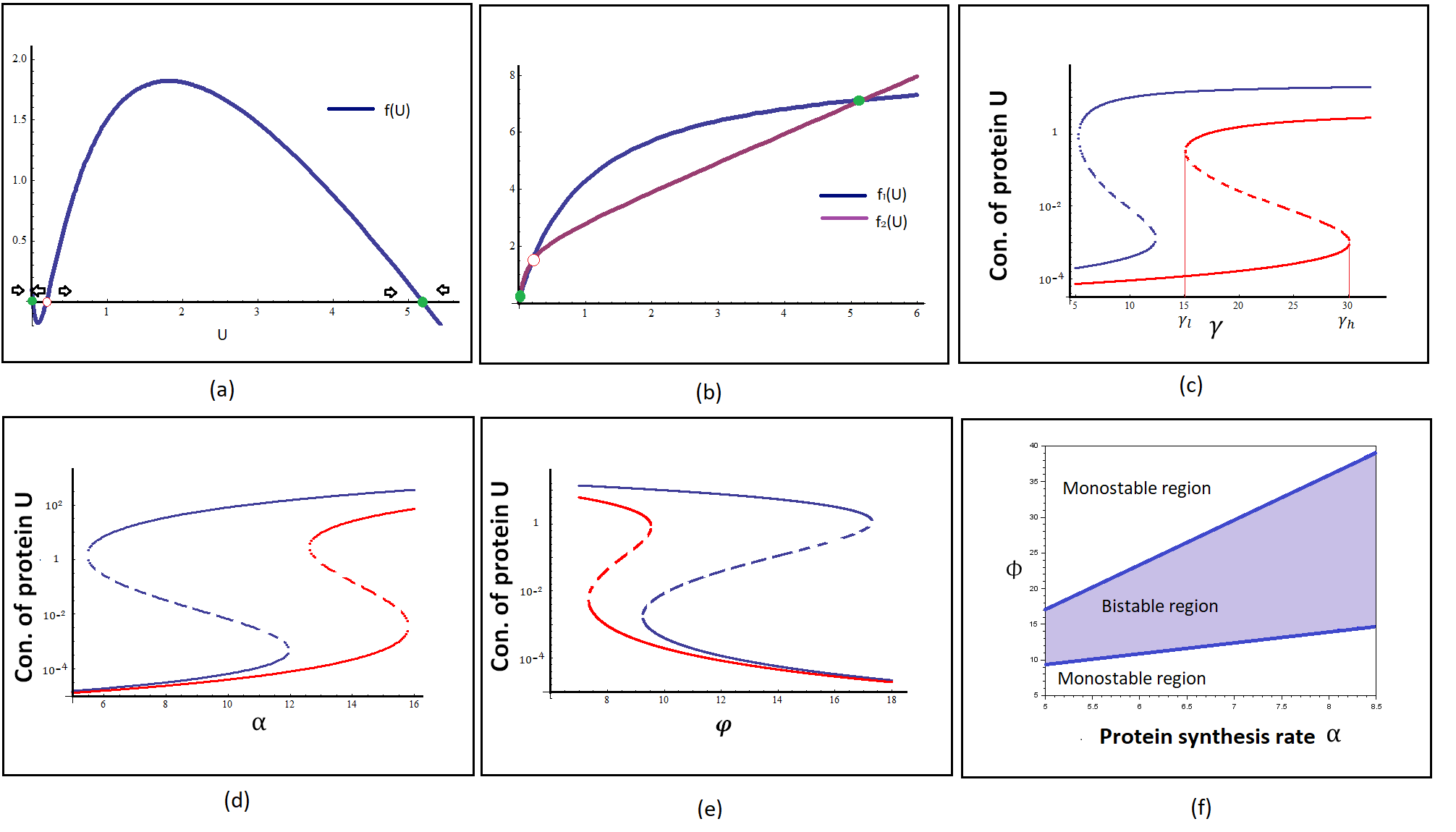}
    \caption{Bifurcation analysis of the model motif. (a) System equation $f(U)$ vs. $U$ plot. $f(U)$ intersects the $U$ axis in $3$ different points, representing $3$ steady states of the system in phase space. Among these $2$ are stable point represented in solid green circle and one is unstable point represented in red hollow dot. Black arrows indicates the direction of flow lines in phase space.  (b) Growth function, $f_1(U)$ and decay function, $f_2(U)$ is plotted against $U$. Two curve intersects in $3$ different points, generating $3$ possible solution of the system. (c) Protein $U$ shows bistability wrt. $\gamma$. Parameter values are $\phi=10$, $\alpha=3$ for the red curve and $\alpha=5$ for the blue curve. (d) Protein $U$ shows bistability wrt. $\alpha$. Parameter values are $\phi = 20, \gamma= 2.5$ for red curve, $\gamma=10$ for the blue curve. (e) Bistability curve of protein $U$ wrt. $\phi$. Parameter values are $\alpha=5$, $ \gamma = 10$ for blue curve, $\gamma=5$ for the red curve. (f) Bifurcation diagram in the plane of parameter $\alpha$, the synthesis rate of protein vs. parameter $\phi$, maximum dilution rate of protein. Bistable region is shown in blue-gray color with a defined blue boundary which separates it from two monostable region.  For (a), (b), (f) $\gamma=10$. For (a), (b) $\alpha=8.5, \phi=20$. For all curves $\delta = 0.1, \beta = 1.15, \Delta_U= 1$. }
    \label{bifurcation}
\end{figure}
\section{Results: Deterministic System}\label{res:bifur}
\subsection{Equilibria \& Stability:}
Biologically significant equilibria for this given system should be non-negative solutions where $\frac{du}{dt}=0$. In other words, if we express Eq. \ref{eqn main} in terms  growth function, $f_1(U)$  and decay function, $f_2(U)$, then,
\begin{equation}\label{eq prt}
     \frac{dU}{dt}=f(U)=f_1(U)-f_2(U)
\end{equation}
where
\begin{equation*}
f_1(U)=\frac{\delta + \alpha\:U}{1+U}\;\;\;\;\;\;\; \text{and} \;\;\;\;\;\;\;\;  f_2(U)=\frac{\phi\:U}{\beta+\gamma\:U}+\Delta_U\:U  
\end{equation*}
then, non-negative equilibrium points will only occur when growth function curve intersects with decay function curve for $U \geq 0$.\\
We started by investigating the steady-state dynamics by exploring the phase trajectories. 
The study of phase space behavior 
in terms of nature, number and relative arrangements of steady states qualitatively signifies what the system can or cannot achieve \cite{strogatz1994nonlinear, hirsch2012differential}.
\\To estimate flow trajectories in phase space, we first observe the dynamical function, $f(U)$ may intersect with the $U=0$ axis for $U \geq 0$ at $3$ different points at most, as shown in Fig. \ref{bifurcation}a.  This refers to the existence of  three biologically feasible  steady states, at the most.  We denote these equilibria as $E^*_L$, $E^*_I$ and $E^*_H$ respectively, where the subscripts stand for low, intermediate and high level of protein concentration.\\
To draw the phase trajectories, it is needed to determine the stability of the equilibria. We use Linear Stability Analysis (LSA) for classifying equilibrium points under small perturbation. In presence of small fluctuation or perturbations, if the trajectories return to $E^*$, then it is characterised as a stable steady state or \textit{attractor}. For a system described by a single ordinary differential equation, ${\dot {x}}= f(x)$ with an equilibrium point ${E^{\star}}$, using Taylor series expansion around $E^*$, we linearize the equation as,
\begin{equation}
f( {x})= f( {E^{\star}})+{\frac{\partial f}{\partial x}\Big|_{ {E^{\star}}}}( {x}- {E^{\star}})
\label{eq:taylor}
\end{equation}
Now, let us consider a small perturbation $\delta x$ from $E^*$; the dynamical system will be expressed by, $f(x)={E^{\star}}+\delta x$. In this scenario, stability is dictated by the gradual growth (or decay) of the perturbation, so that system evolves towards (or away from) the steady state ${E^{\star}}$, which classifies it as stable (or unstable) solution. So, to study the behavior of $\delta   x$ with time, we take a time derivative, to find that,\\
\begin{equation}
\label{eq:delx}
{\delta \dot{x}} = {\dot{x}}\\
= {f}({x}),
\end{equation}
as $ {E^{\star}}$ is a constant. Comparing Eq. \ref{eq:taylor} and \ref{eq:delx}, we get
\begin{equation}
 \delta {{\dot {x}}}=\frac{\partial f}{\partial x}\Big|_{ {E^{\star}}}\delta {x},
\end{equation}
For  ${E^\star}$ to be \textit{stable},  $\frac{\partial f}{\partial x}\Big|_{ {E^{\star}}}=f'(E^*)$ should be \textit{negative}. The magnitude of  $f'(E^*)$  evaluates the strength of the attractor.
Using LSA, we classify and mark stable fixed points by green solid circle and  the unstable fixed point by red hollow circle (Fig. \ref{bifurcation}(a) and (b)). We find three regimes of stability:
\begin{itemize}
    \item \textit{Regime 1:}, where only $E^*_L$ exists, and it is stable,
    \item \textit{Regime 2:}, where only $E^*_H$ exists, and it is stable, and
    \item \textit{Regime 3:}, where on \textit{all} equilibrium points exist, with $E^*_L$ and $E^*_H$ being stable, while $E^*_I$ being unstable.
\end{itemize}
We show the phase trajectories using arrows just above $U=0$ axis in Fig. \ref{bifurcation}(a). We can also arrive at the same conclusion about stability by comparing the slopes of growth function, $f_1(U)$ and decay function, $f_2(U)$, as shown in Fig. \ref{bifurcation}(b). It shows depending upon  threshold slopes of both these functions, the curves may intersect each other in $3$ different points; the stability can also be estimated from this figure by graphical approach \cite{strogatz1994nonlinear}. Regime 3 depicts a scenario of bistability; thus we proceed to further analyze the bifurcation. 
\subsection{Bifurcation analysis:}
We observe that the system undergoes a saddle-node bifurcation as certain thresholds are crossed. In Fig. \ref{bifurcation}(c) $\rightarrow$ \ref{bifurcation}(e), we have shown the bistable behavior of protein $U$ wrt. different system parameters ($\gamma$, $\alpha$, $\phi$ respectively). Fig. \ref{bifurcation}(c) shows for a range of $\gamma_l$ to $\gamma_h$ the system has a choice between $E^*_L$ and $E^*_H$. The two states are separated by the unstable $E^*_I$. This is the region of bistability, in which the system is also capable of exhibiting history-dependent response. The solid lines represent the stable state and the dotted lines are representing the unstable states. We also observe for higher values of $\alpha$ the region of bistability shrinks. Similar responses can be observed for other parameters in Fig. \ref{bifurcation}(d) and (e). \\
We also have further done the phase space plot in the plane of parameter $\alpha$, the synthesis rate of protein vs. parameter $\phi$, maximum dilution rate of protein. Bistable region is shown in blue-gray color with a defined blue boundary which separates it from two monostable region in Fig. \ref{bifurcation}(f). Here, it is important to note that phase space of a dynamical system is an abstract space, and here the value of a specific state variable is represented by the dimensions used.

\section{Results: Reaction diffusion model}\label{res diff}
We now proceed to consider the spatio-temporal evolution of the system in presence of diffusion. Diffusion is a fundamental mechanism in biological systems. From drug dispersion to allocation of fundamental agents in the body, diffusion plays essential role. In the dynamic cell environment, diffusion allows the interaction between macro molecules, such as substrates which needed to find enzymes. Similar for  transcription factors and binding sites on the DNA, membrane proteins and membrane etc. Different models on protein diffusion includes diffusion within plasma membrane (fluid mosaic model) as well as diffusion outside plasma membrane with the help of some delivery system \cite{jacobson1995revisiting, beznoussenko2014transport}. 
\subsection{Spatially Explicit Model:}
In Eq. \ref{eqn main}, now we will include one dimensional diffusion in the above system, considering a group of cells, arranged in form of a linear, one-dimensional chain/string. Each cell contains a single genetic circuit looking exactly like Fig. \ref{motif}. and diffusible protein molecules of $U$ are allowed to diffuse through the whole one dimensional space. The mathematical form of this system is represented by Eq. \ref{eqn 1d}:
\begin{equation}\label{eqn 1d}
    \frac{\partial U\:(x,t)}{\partial t}= \frac{\delta + \alpha\:U}{1+U} -\frac{\phi\:U}{\beta+\gamma\:U}-\Delta_U\:U + D_U\:\frac{\partial^2U}{\partial x^2}
\end{equation}
\\where, $x$ represents the position in space and $D_U$ represents the rate of diffusion of protein $U$ through the one dimensional chain of cells.\\
Next, we simulate a two dimensional, mono-layer/film made of cells and the protein can diffuse through the plane. We incorporate 2D diffusion into Eq. \ref{eqn main}, and achieve the spatially explicit dynamical system given by Eq. \ref{2D}:
 \begin{eqnarray}\label{eqn 2d}
    \frac{\partial U\:(x,y,t)}{\partial t}= &=& \frac{\delta + \alpha\:U}{1+U} -\frac{\phi\:U}{k+\gamma\:U}-\Delta_U\:U+D_U^x\;\frac{\partial^2U}{\partial x^2}+D_U^y\;\frac{\partial^2U}{\partial y^2} \nonumber \\
    & \equiv & f(u) +D_U^x\;\frac{\partial^2U}{\partial x^2}+D_U^y\;\frac{\partial^2U}{\partial y^2}
    \label{2D}
\end{eqnarray}
where, $(x,y)$ represent the position of the cell, and the diffusion rate constants for $x$ and $y$ direction are given by  $D_U^x$ and $D_U^y$. This diffusion rate constants denote the diffusivity strength of the protein in a direction-dependent manner, representing how fast or slow it can diffuse through cell membrane and how the local information of concentration is getting distributed globally.

\subsection{Stability of the system:}

In Eq. \ref{2D}, $f(u)$ is the reaction term of the dynamical system, $\frac{\partial ^2 U}{\partial x^2}$ and $\frac{\partial ^2 U}{\partial y^2}$ capture the spatial fluxes of the concentration of the protein $U$ in $x$ and $y$ directions in space. To understand the stability of the system, let us linearize $f(u)$ about $u_0$, the stable fixed point of the dynamical system in absence of diffusion, gives
\begin{equation}
f(u) \approx f(u_0) + f_u \delta_u + ...
\end{equation}
where 
\begin{equation}
f_u = \left. \frac{\partial f}{\partial u} \right\vert_{(u_0)} = -1 + \frac{\alpha}{1 + u_0} - \frac{\delta + \alpha u_0}{(1 + u_0)^2} + \frac{\gamma \phi u_0}{(b + \gamma u_0)^2} - \frac{\phi}{b + \gamma u_0}
\end{equation}
is the first derivative of the reaction term and $\delta u(x,y,t) = u(x,y,t) - u_0 $ is the perturbation from the homogeneous steady state. Thus, the linearlized system,

\begin{equation}
\frac{d}{dt}(\delta u) = f_u \delta u + D_U^x\;\frac{\partial^2U}{\partial x^2} + D_U^y\;\frac{\partial^2U}{\partial y^2}
\end{equation}
Taking Fourier transform, we obtain the following ordinary differential equation
\begin{equation} \label{ODE}
\frac{d}{dt}(\widehat{\delta u}) = -k^2(D_U^x+D_U^y)\widehat{\delta u} + f_u \widehat{\delta u}
\end{equation}
where $\overrightarrow{k}=(k_x,k_y)$
Let us assume the solution of this linear system of the following harmonic form (keeping the meaning same, we drop the `hat' for convenience; henceforth $\delta u$ refers to the Fourier transformed variable.)
\begin{equation} \label{Harmonic_soln}
    \delta u = \delta u_0 e^{-i(\overrightarrow{k_x}.\overrightarrow{x}+\overrightarrow{k_y}.\overrightarrow{y})+\lambda t}
\end{equation}
where $\overrightarrow{k} = (k_x,k_y)$ is the vector of wavenumbers of two-dimensional spatial system and $\lambda$ is equivalent to eigenvalue in multidimensional system. Here, this can also be considered as \textit{inverse characteristics time}, $\tau^{-1}$, which gives an estimation of time required for relapse to the steady state. Differentiating Eq. \ref{Harmonic_soln}, we obtain 
\begin{equation}
\frac{d}{d t} \delta u = \tau^{-1} \delta u
\end{equation}
Comparing Equations \ref{ODE} and \ref{Harmonic_soln}, we see that $\lambda$ is given by
\begin{equation} \label{Eq_eigen}
    \tau^{-1} = -k^2(D_U^x+D_U^y) + f_u
\end{equation}
For diffusion in one dimension, the above expression reduces to $\tau^{-1} = -k^2 D_U^x + f_u$.

\begin{figure}
    \centering
    \includegraphics[width=\textwidth]{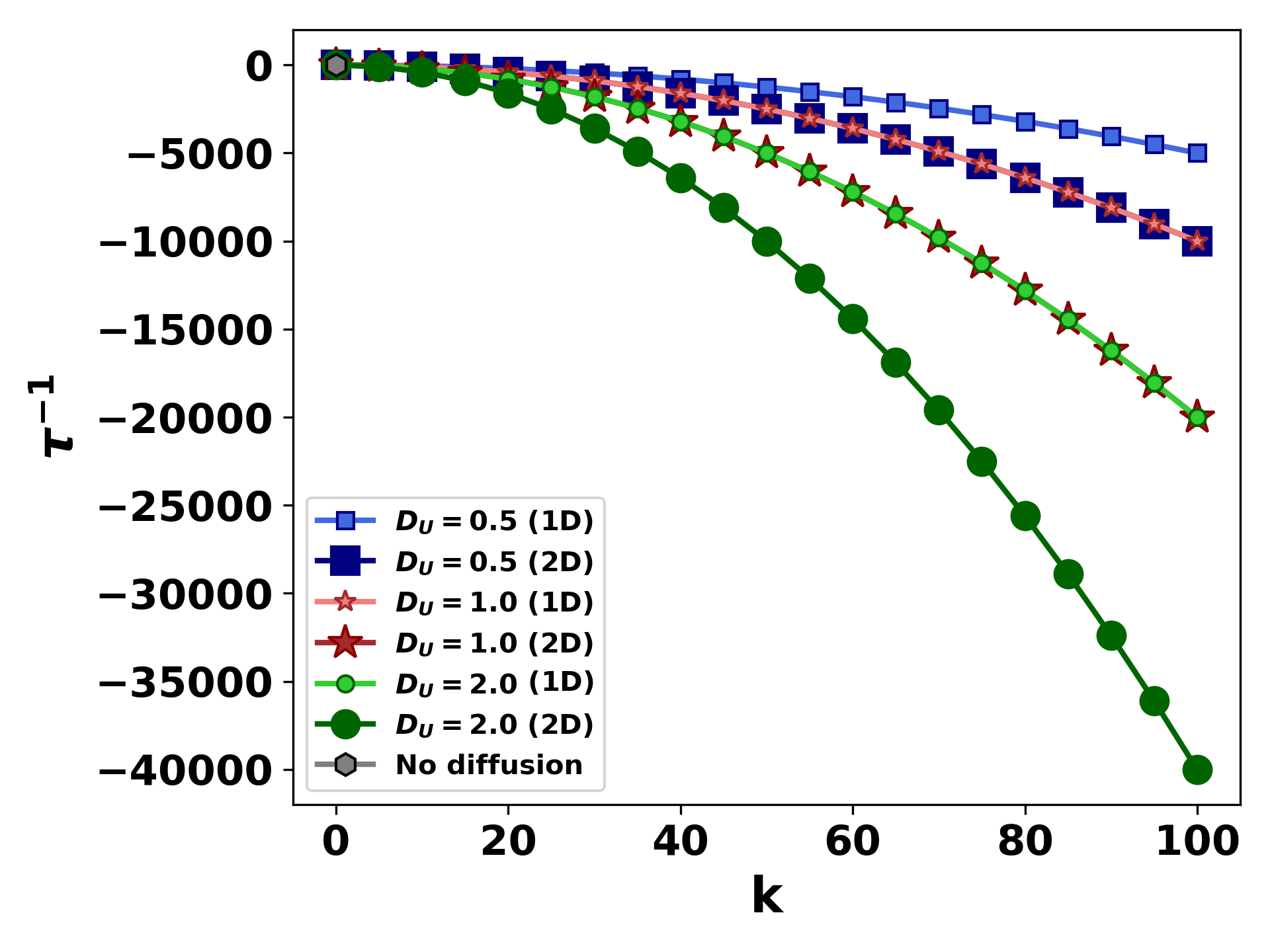}
    \caption{Illustration of the nature of the inverse characteristics time, $\tau^{-1}$ (evaluated at one of the steady states) of the system as a function of wave number $k$ for three different values of the diffusion coefficients $D_U$ in two dimensions. Corresponding lighter shades (with same symbols but smaller in size) are for the $\tau^{-1}$ when spatial diffusion is considered in one dimension. For $k=0$, the system has no diffusion (denoted by a single point in grey). Qualitatively similar result is found in the study of the other steady state (not shown here).}
    \label{eigen}
\end{figure}

The variation of the inverse characteristics time, $\tau^{-1}$ (given in Eq. \ref{Eq_eigen}) is illustrated as a function of the wavenumber $k$ in Fig. \ref{eigen}. The nature of the stability of the dynamical system can be evaluated from the sign of $\tau^{-1}$. Analytical solutions reveal negative $\tau^{-1}$, thus indicating the steadyness of the patterns. With the increase in wavenumber $k$, $\tau^{-1}$ become more negative. For $k = 0$, the system reduces to cases of diffusion-free pure reaction type dynamical system. In all the cases, $\tau^{-1}$ becomes negative more rapidly for diffusion in 2D than that in 1D. This suggests that the patterns homogenizes much quicker in 2D. We notice that the results for $D_U=2.0$ in 1D matches with that of $D_U=1.0$ in 2D and $D_U=1.0$ in 1D matches with that of $D_U=0.5$. This is simply because, in the Fourier space, the first term in the expression for $\tau^{-1}$ has a multiplicative factor of Diffusion coefficient.
\subsection{Effect of Initial Conditions:}
Next, to study the diffusion pattern in our current host-circuit  interaction mediated model we consider different initial conditions. As a consequence of different ongoing process in the cell, a combined signal of positional information is generated on the circuit of interest and it is very interesting to observe how this information is generated as well as how the circuit reacts to this signalling cues. Some of the recent works have established that this positional information can be carried by some diffusing bio-chemical morphogen \cite{gierer1972theory, tata2000autoinduction}. Depending upon the concentration of a local ligands, activation of a target gene, as well as the gene expression  is regulated. Another model of A. Turing suggested existence of chemical gradient in biological systems which pre-patterns the system followed by cell differentiation and further pattern formation.  Considering these biological variations, we include different initial conditions to study our concerned motif. We also include random initialisation, as most of the times the spatial information is ramdomised within a certain scale for a bacterial population. It has been also established that, in some orgamisms, like \textit{Hydra}, pattern formation results from an initial mass of cells without any precise positional information \cite{muller1997model, hobmayer2000wnt}. 
\\Detailed results of spatio-temporal pattern formation in one dimension and two dimension with different initial conditions are mentioned below.

\subsubsection{Random initialization:}
\begin{figure}
    \centering
    \includegraphics[width=\textwidth]{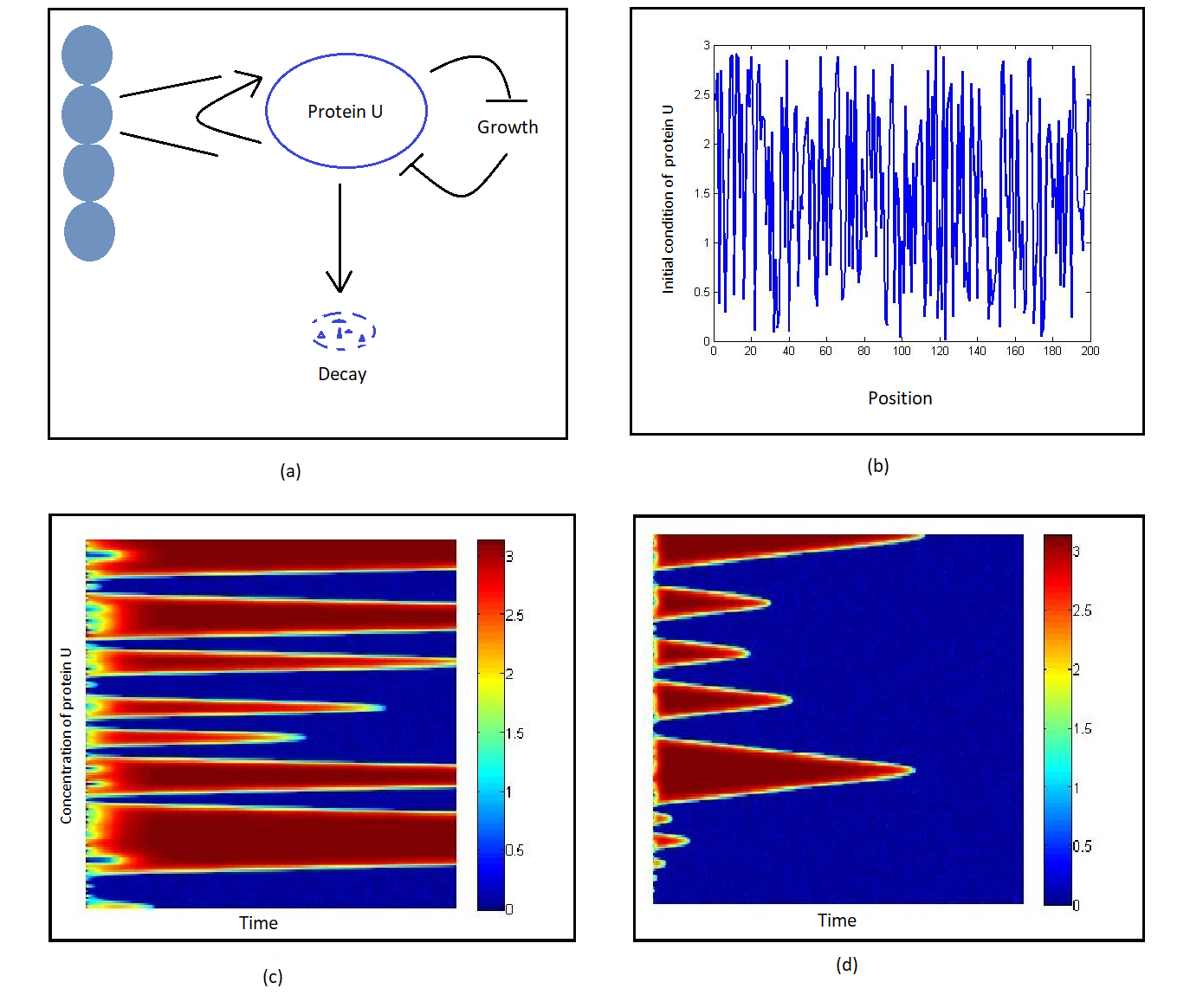}
    \caption{One dimensional diffusion of the motif with random initial  condition. (a) Schematic diagram of a chain of cell each containing the motif. (b) A schematic distribution plot of initial condition of protein $U$, wrt. position. (c) Diffusion pattern after time 50. (d) Diffusion pattern after time 250. Diffusion coefficient $D_U=1$, Parameter values are $\epsilon=3$, $\alpha=8.5$, $\gamma=1.2$, $\phi=4$, $\beta=0.01$, $\delta= 0.03$, $\Delta_U=1$.Red and blue denote high and low expression levels of protein $U$. As time evolves, the system traverses towards a homogeneous state of the low expression level of the protein.}
    \label{1d random }
\end{figure}

Now, we focus as the condition described in Eq. \ref{eqn 1d}, and as shown in Fig. \ref{1d random }(a), where we allow diffusion in one dimensional chain of cells, each containing a motif of interest. The position, $x$ has been discretized as $x_i$, where $i \in \{0, 200\}$ presenting 200 cells. The time has also been discretized with a time-step $\Delta t=0.001$, and no flux boundary condition has been considered. We consider the initial condition as shown in Fig. \ref{1d random }(b), where, $U(x_i,0)$ is given by uniform random distribution $\xi(0,1)$, scaled by a factor of $\epsilon$.
Fig. \ref{1d random }(c). and Fig. \ref{1d random }(d). represents the time variation of the diffusion pattern. Diffusion constant $D_U$ is fixed at the value of $1$. As time passes, though started from a random initial distribution, reaction-diffusion causes the system to show patterns with defined boundaries where two state with drastic different value of protein concentration (low and high) exists side by side, the local concentration spreads spatially, influencing its neighbouring cells, producing islands of either high synthesis region (shown in red color in the figures) or of low synthesis region (shown in blue color). Further we can notice that, as the time passes red islands are getting smaller and then vanishes, and the system has only low synthesis state, indicated by an overall spread of blue color (comparing Fig. \ref{1d random }(c). taken after time $t=50$, and Fig. \ref{1d random }(d). taken after time $t=250$.). 
\subsubsection{Initialization with Exponential Gradient:}
\begin{figure}
    \centering
    \includegraphics[width=\textwidth]{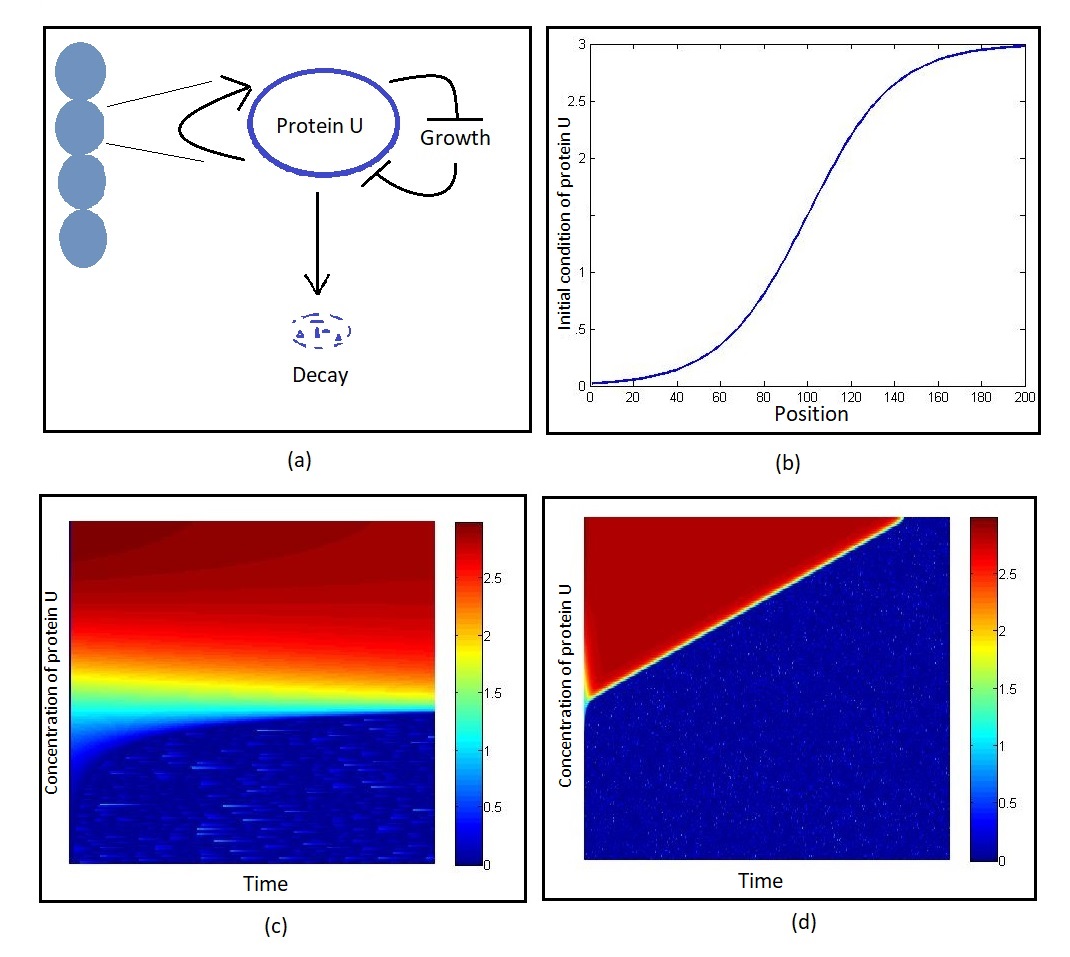}
    \caption{One dimensional diffusion of the motif with exponentially increasing initial  condition.  $k_1=3,\; k_2=0.05$ and $L_{half}=100$. (a) Schematic diagram of a chain of cell each containing the motif. (b) A schematic distribution plot of initial condition of protein $U$, wrt. position. (c) Diffusion pattern after time 2. (d) Diffusion pattern after time 250. Diffusion coefficient $D_U=2$, Parameter values are $\alpha=8.3$, $\gamma=1.4$, $\phi=4$, $\beta=0.01$, $\delta= 0.03$, $\Delta_U=1$. Red and blue denote high and low expression levels of protein $U$.}
    \label{1d exp}
\end{figure}

Next, we consider the diffusion in one dimensional chain of cells with an exponentially increasing concentration of initial condition as shown in Fig. \ref{1d exp}(b). This type of diffusion pattern is very common and similar can be seen in toggle switch spatial one dimensional diffusion, reported in \cite{roy2022spatiotemporal}.  Initial conditions are taken such that the mid-value of concentration crosses the position line at $L_{half}$, i.e. at the mid-point of the considered 200$\times$1 grid of cells as mentioned before. The mathematical form of initial condition is given below.
\begin{equation*}
U_{initial}=\frac{k_1}{1+exp(k_2*(L_{half}-x_i))}    
\end{equation*}
We find from Fig. \ref{1d exp}(c), and Fig. \ref{1d exp}(d) that within very small time the spatial information distributed in nearly half of the considered total grid of cells as a low synthesis region with blue shade and other half of the grid as high synthesis region with red shade. As time passes, the low synthesis region spreads over and the red island of high synthesis region is getting smaller and finally vanishes in Fig. \ref{1d exp}(d). Same boundary conditions and time steps are used as before. However, to avoid numerical artifacts, different values of $\Delta t$ were considered, which gave qualitatively similar results.

\subsection{Stochastic Periodic Initialization:}
\begin{figure}
    \centering
    \includegraphics[width=\textwidth]{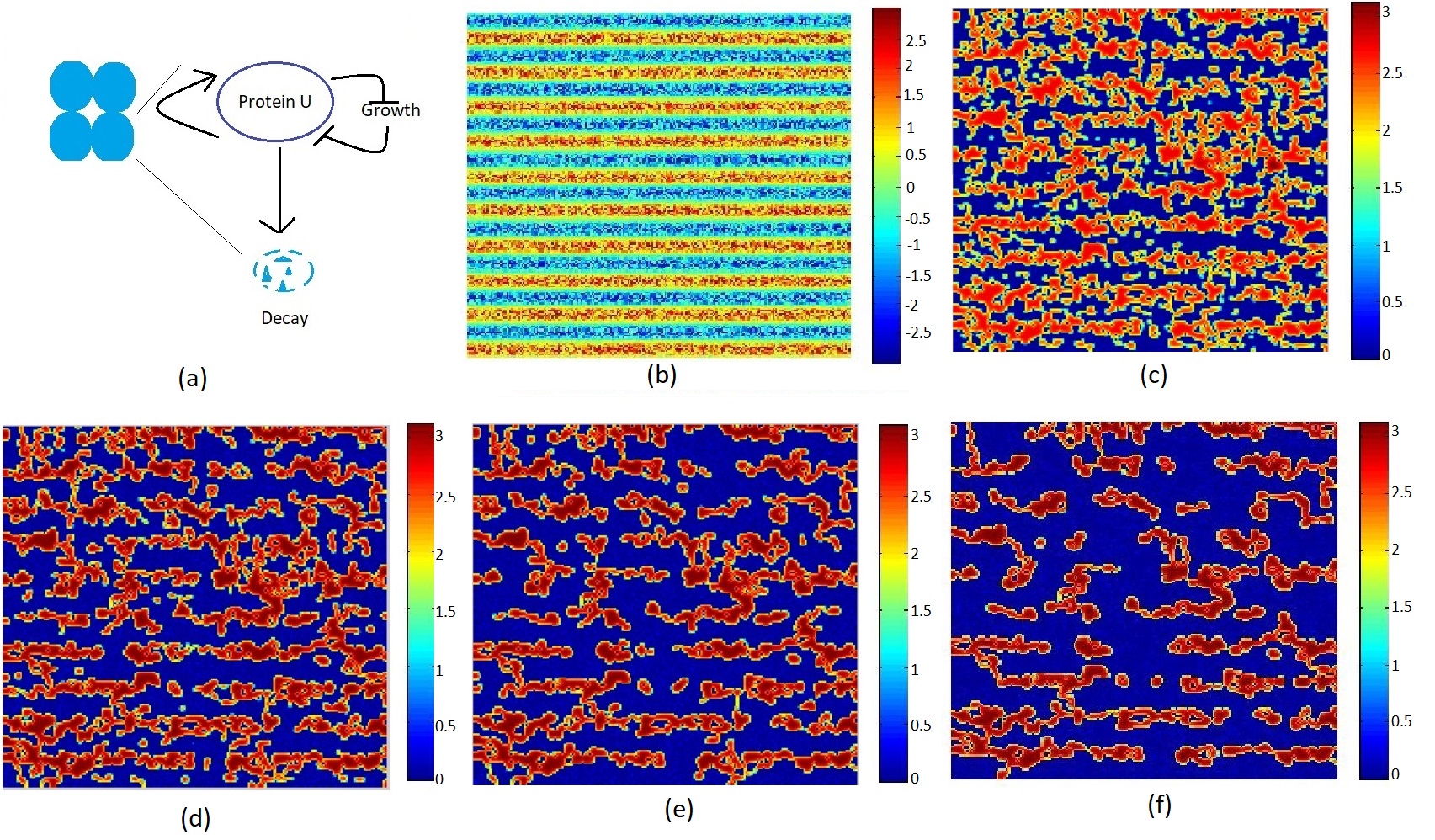}
    \caption{Diffusion in two dimension with randomly distributed periodic function, for $k_2$ = $10$. (a) Schematic figure illustrating a two-dimensional sheet of cells, each cell having the model motif with protein-molecules diffusing across the two-dimensional sheet. (b) Distribution of initial condition. (c) - (f) shows diffusion pattern after different time instants (c) 20. (d) 50. (e) 100. (f) 400. Parameter values are $\epsilon=3$, $\alpha=8.5, \gamma=1.2, \phi=4, \beta=0.01, \delta=0.03, \Delta_U=1, D_U=0.2$ }
    \label{2dransin}
\end{figure}

Now, we move to a two dimensional sheet of 40000 cells arranged in a $200\times200$ spatial arena (schematic diagram in Fig. \ref{2dransin}(a)). Position of each cell is denoted by $(x_i,\; y_i), \;\; i \in \{1,\;200\}$ as each cell contains one motif, that expresses diffusible protein molecule $U$. In this analysis, we consider isotropic diffusion, i.e., $D_U^x = D_U^y = D_U$. In this set-up, we study the system under various initial conditions, dependent on the discrete position in space $(x_i,\; y_i)$. While we will elaborate on the random initial condition in the upcoming section with detailed quantitative treatment, we report interesting response we observe for a sinusoidal initial condition with added stochasticity in this section (Fig. \ref{2dransin}(b)).  Spatio-temporal evolution of the pattern of the motif is shown in Fig. \ref{2dransin}(c), (d), (e), (f). The mathematical form of initial condition is given below. 
\begin{equation*}
    U_{initial}=\epsilon \xi(0,1)\:sin \frac{\pi\:x_i}{k_2}
\end{equation*}
where, $\xi(0,1)$ represents uniform random distribution, with a scaling factor $\epsilon$.  We can see some beautiful stripe like pattern in output which remains for considerably long time and eventually dies out. Stripe pattern is one of the most abundantly found pattern in living systems (including zebra, tiger etc. for animals, leaf pattern in calathea ornata, calathea majestica etc.). But these refers to stable patterns while the pattern in our model is transient, and after existing for a long time, the pattern dies out. This shows the significance of intermediate levels of transient phenotypic heterogeneity arising due to initial conditions.
The time evolution of the pattern for other biologically relevant initial conditions like exponentially growing or decaying gradient have been reported in Appendix (Fig. \ref{2dtan}). All these patterns persisted for substantially long time.
\begin{figure}
   \centering
    \includegraphics[height=0.82\textheight, width=\textwidth]{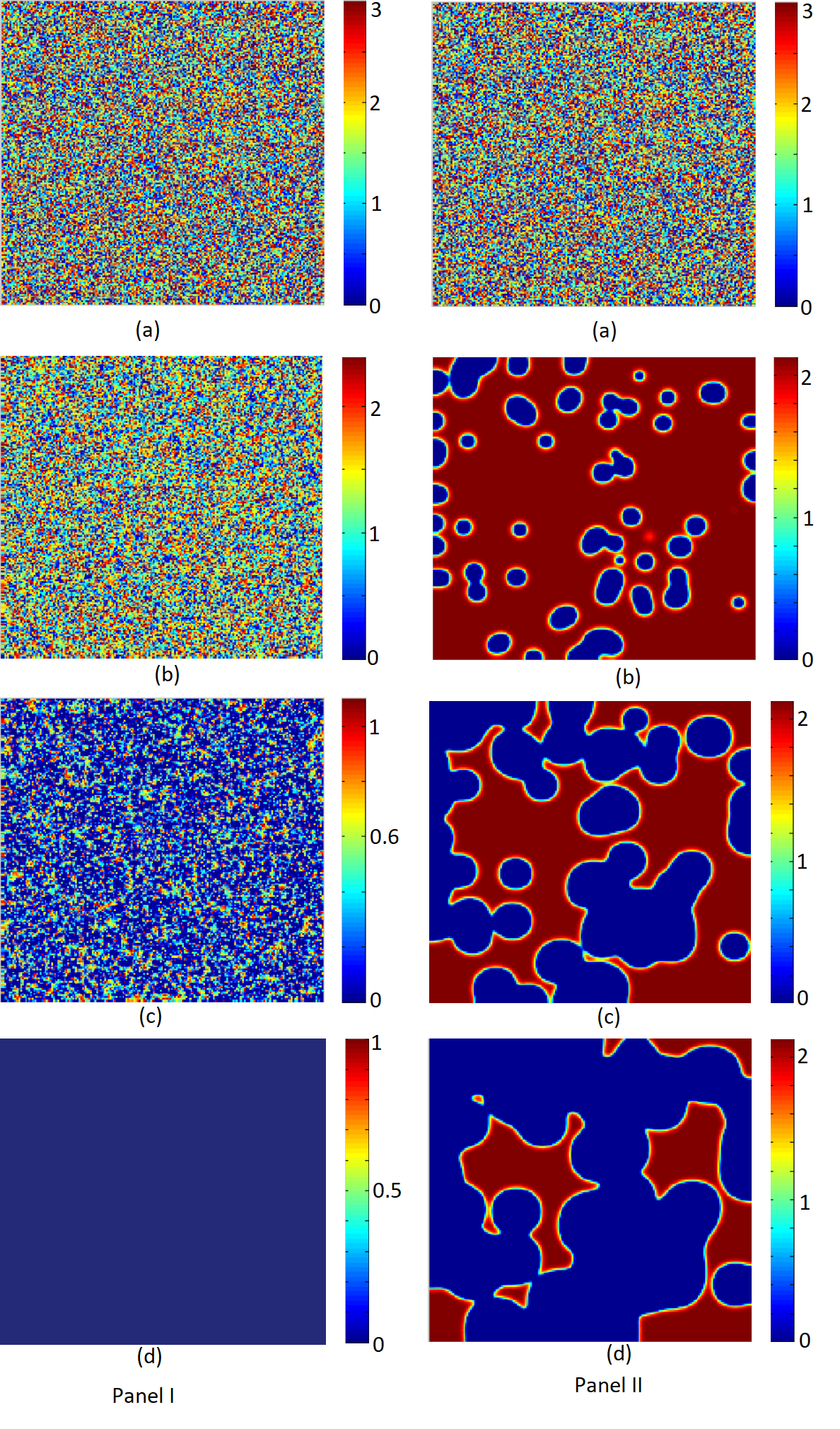}
    \caption{Diffusion in two dimension in absence (Panel I, $\gamma=0$) and presence (Panel II, $\gamma=4$) of host-circuit interaction with random initial condition.(a) An initial distribution plot of protein $U$ in a random way. Pattern after different time instants, for panel I, (b) $0.0007$. (c) 0.002. (d) 0.0035. For panel II, (b) $25$. (c) $60$. (d) $85$. Rest of the parameter values are $\alpha=9.6$, $\phi=20$, $\beta=1.15$, $\delta= 0.01$, $\Delta_U=1, D_U=0.5$ for both panel I and panel II.}
    \label{hstckt}
\end{figure}
\section{Quantitative Analysis of Spatio-temporal pattern formation:}\label{quanti}
Our prime concern is about transient pattern formation by host-circuit interaction in gene expression dynamics, and explore the quantitative differences with the scenario without any host coupling. In order to quantify the major observations, we perform following analyses.
\subsection{Essentiality of Host-circuit interaction for Pattern Formation:}
To verify the essentiality of host-circuit interplay for the observed dynamics, we consider a two dimensional array of 200$\times$200 cells each containing one motif with diffusing molecules. We consider a randomly distributed initial condition, 
\begin{equation*}
    U(x_i, y_i, 0)=\epsilon \xi (0,1)
\end{equation*}
defined by a scaled uniform random distribution $\xi(0,1)$, and no flux boundary condition. We study the system when nonlinear host-circuit interaction is OFF (with the $\gamma$ value $0$, as shown in Fig. \ref{hstckt}, panel I ) and the interaction is ON (putting $\gamma =4$ shown in \ref{hstckt}, panel II ). In presence of host-circuit interaction, from the initial condition as shown in Fig. \ref{hstckt}(a), as time passes, we find some beautiful pattern in two dimension where the low synthesis state and high synthesis states  coexist spatially as shown in panel II of Fig. \ref{hstckt}(b),(c),(d). High synthesis state is shown in color red and low synthesis state is shown in color blue. The change in patch size is also clearly visible. This patterns are transient in nature and remains for some time, the system gradually converges to its low synthesis stable state after considerable amount of time. However in panel I, Fig. \ref{hstckt}, we can see the system quickly converges to its monostable low synthesis state, no visible pattern is forming (panel I, Fig. \ref{hstckt}(a)-(d). Thus, the comparison between Fig. \ref{hstckt} panel I and panel II clearly signifies that the spatio-temporal pattern arising in the motif is completely regulated by the host-circuit interaction, an emergent phenomena and not the system's inherent property. 
\begin{figure}
    \centering
    \includegraphics[width=\textwidth]{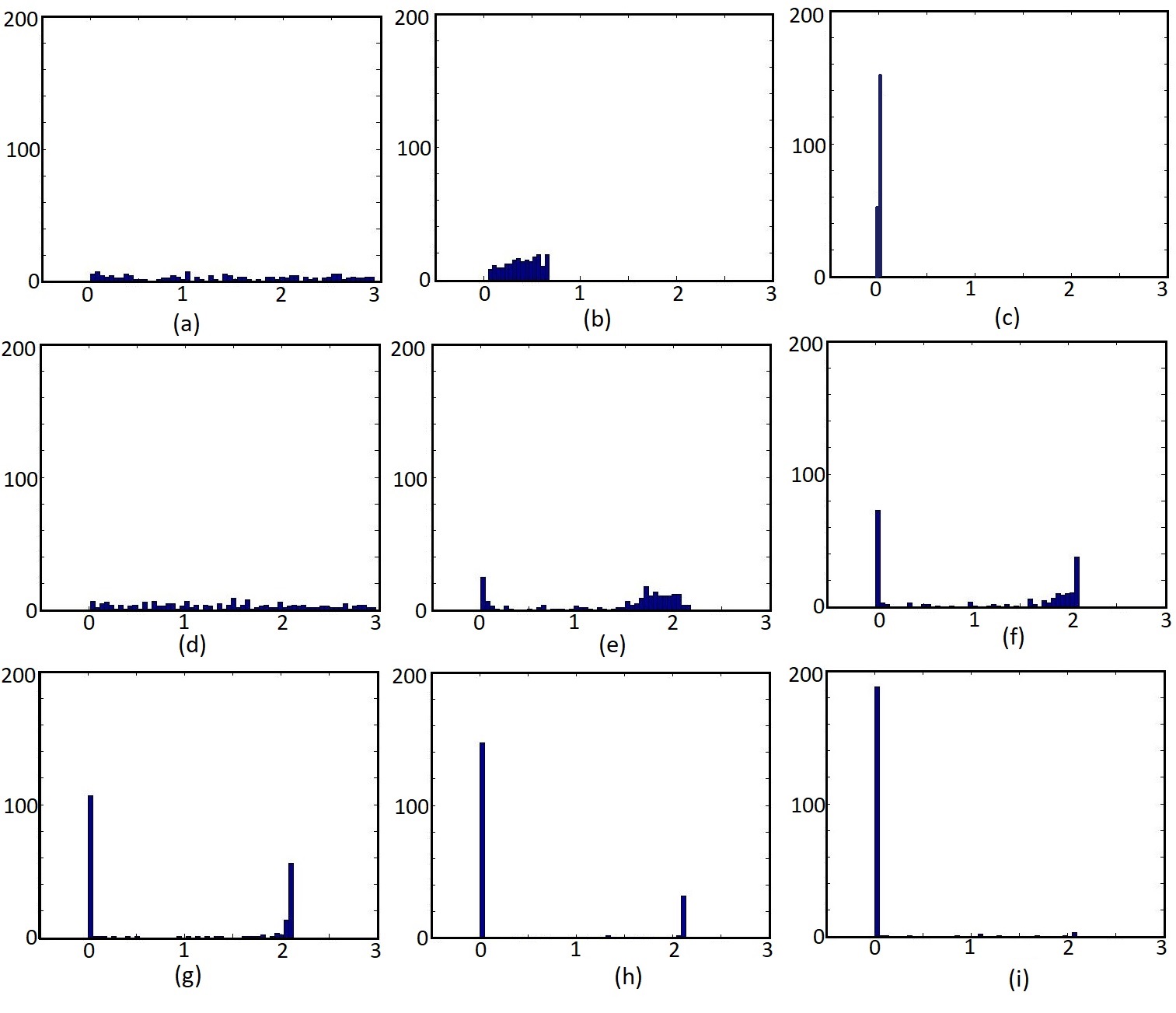}
    \caption{Histogram plot for the distribution of cells and preferred steady states without and with host-circuit interaction. x axis represents the concentration of cell, y axis represents the number of cell with that particular concentration. (a), (b), (c) are for without host-circuit interaction ($\gamma=0$). (a) Initial distribution of cells. Histogram plot after different time instant. (b) 0.1. (c) 0.5. (d) $\rightarrow$ (i) are for host-circuit interaction model ($\gamma=4$). (d) Initial distribution of cells. Histogram plots after different time instants. (e) 5. (f) 10. (g)  20. (h) 30. (i) 40. parameter values are $\alpha =9.6, \phi =20, \delta=0.01, \beta= 1.15, D_U=0.5, \Delta_U=1 $.}   
    \label{histogram}
\end{figure}
\subsection{Histogram plots in presence \& absence of host-circuit interactions:}
For further quantification, in the one dimensional cellular array, we draw the histogram plots to emphasise the distribution of cells in different steady state in presence of diffusion, with and without host-circuit interaction. We start with initially distributed set of cells which have concentration ranging from $0 \rightarrow U_{max}$ at time $t=0$. We divide the range $0 \rightarrow U_{max}$ in bins of size 0.05. \\
In absence of host-circuit interaction (putting $\gamma =0$), diffusion causes the system to converge strictly to its monostable steady state (Fig. \ref{histogram}(a),(b),(c)) as time passes. Now, in presence of host-circuit interaction with the initially distributed cells of concentration $0 \rightarrow U_{max}$, the diffusion in the system  clearly signifies two stable states as time passes. A bimodal distribution is clearly seen where more cells are accommodating in these two modes for an intermediate time interval. The increase in bar length on the two particular position, indicating most of the cells are preferring these two concentration (thus two stable steady state Fig. \ref{histogram}(e)-(h)). Though, we can see further the bar length in high concentration stable state decreases gradually and the cells are converging to low concentration stable state as time passes (Fig. \ref{histogram}(g)-(i)). From this distribution plot, a further comment on transient pattern formation can be done as we can see the bistable nature is only carried out for a range of intermediate time in the system, so the pattern can be seen for an intermediate interval which eventually converges to a single steady state. 
\begin{figure}
    \centering
    \includegraphics[width=\textwidth]{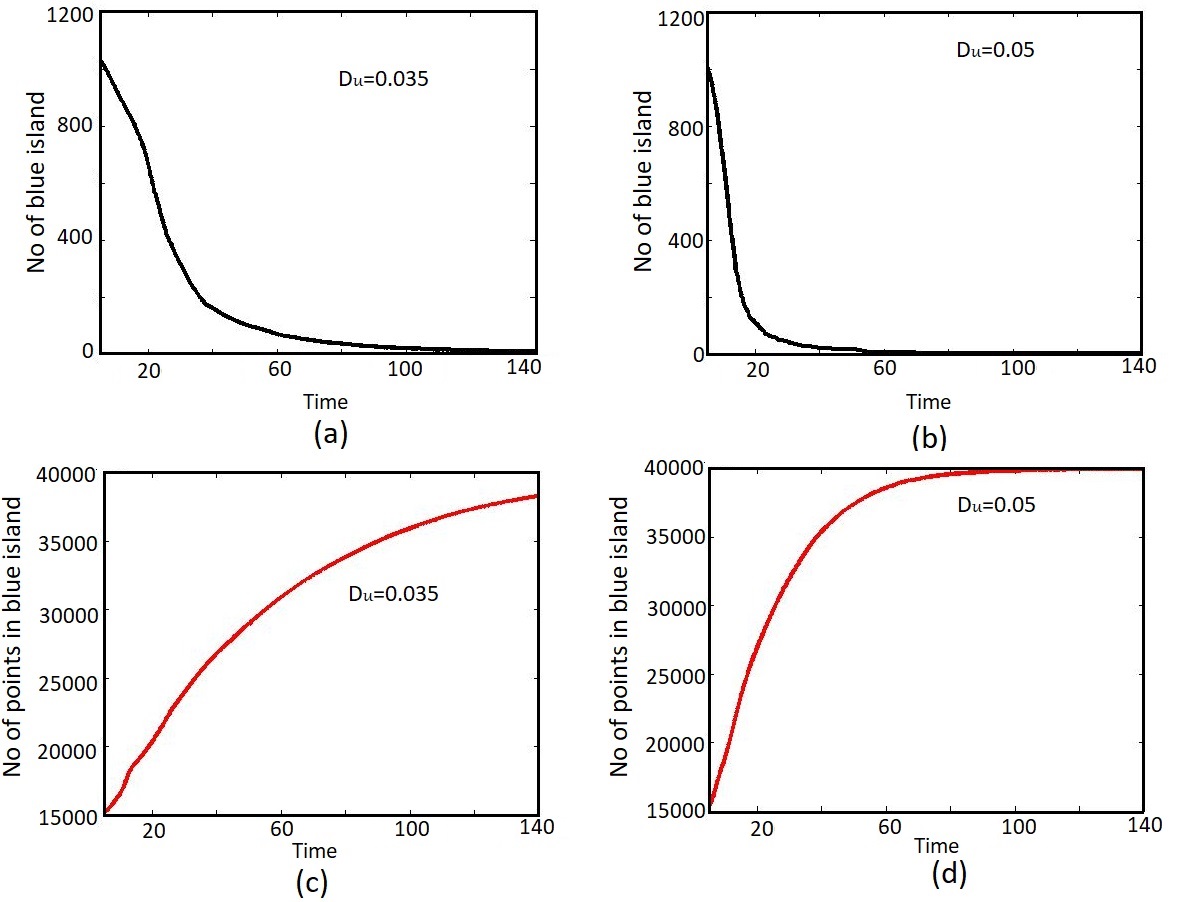}
    \caption{Comparative study for the effects of diffusion coefficient in a randomly distributed two dimensional array of $200 \times 200$ cells, after time 140.  (a) and (c) represents the no of blue islands and the number of point in blue islands respectively for diffusion coefficient $D_U=$0.035. (b), (d) are same for diffusion coefficient $D_U=$ 0.05. Rest of the parameter values are $\alpha =9.6, \phi =20, \delta=0.01, \beta= 1.15, \gamma=4, \Delta_U=1. $}
    \label{dif}
\end{figure}
\subsection{Connected Component Analysis \& Patch size variation:}
Diffusion constant or diffusivity is the measure of how slow or fast the information, in terms of protein concentration diffusing from its local cell to neighbouring global cell. To analyze the spatio-temporal variation of the resultant patterns on every time step, we perform  connected component analysis on the discretized space. The study is performed for similar conditions and parameter values of the dynamic shown in Panel II of Fig. \ref{hstckt}. To be precise, first, we assume the domain of the diffusion process $(x_i,y_i,t_i)\in \mathbb{Z}^3$ and  perform a thresholding operation to estimate $\hat{U}\:(x_i,y_i,t_i)$ such that $\hat{U}\:(x_i,y_i,t_i)$ is one (zero) if $U\:(x_i,y_i,t_i)$ is greater (less) than a threshold $\theta_b$. The threshold $\theta_b$ is decided as $\frac{1}{2}(\mathcal{U}^{max}+\mathcal{U}^{min})$, where $\mathcal{U}^{max}$ and $\mathcal{U}^{min}$ are the maximum and minimum values of $U,\: \forall x_i,y_i,t_i$, respectively. At time $t_i$, we perform the connected component analysis \cite{he2009fast} to count the number of blue islands considering $\hat{U}\:(x_i,y_i,t_i)=0$ as the object points. At any time instance $t_i$, we estimate the total number of blue points by counting the points where $\hat{U}\:(x_i,y_i,t_i)=0$. For comparison, two different diffusion coefficient values are considered and the variation is reported. Fig. \ref{dif}(a),(b) shows that the number of blue island decreasing constantly while Fig. \ref{dif}(c),(d) representing the number of points in blue island has increasing trend wrt. time. Each blue point represents a cell with low synthesis state, simultaneous increase in number of blue point in island and decrease in number of blue island actually means the red color (high synthesis states) are converting to  blue color (low synthesis states). Thus the proportion of low synthesis state is getting increased wrt. high synthesis states, small blue islands are joining together to make large islands.  So the number of blue island is decreasing but total number of blue points in the island is increasing with time.  Further the effect of diffusion coefficient on the stability of the system is very clear form  Fig. \ref{dif}.  For diffusion coefficient 0.035 (Fig. \ref{dif}(a)) we can see a fall in number of blue island and it almost reaches zero as shown in figure. Comparing Fig. \ref{dif}(b) for diffusion coefficient 0.05 we can see the fall is sharper.  As the decrease in number of blue island comes with its consequence of increase in number of points in blue islands, we can see in Fig. \ref{dif}(c),(d) the trend is less sharper in Fig. \ref{dif}(c), compared to \ref{dif}(d).  This quantifies pace of the homogenization in presence of different diffusion constants. 

\section{Discussions}\label{diss}
The emergent regulatory feedback we consider in this work, is a consequence of the growth burden of host, and is capable of introducing bistability in a monostable circuit. As bistability is a major regulator of nongenetic heterogeneity in cell population, this may have significant role to play in natural dynamics in presence of resource limitation. In this work, we focus on the rarely explored area of the effects of spatial diffusion and pattern formation in presence of host-circuit coupling. Spatial patterns can act as a significant controller for development of diversity in cell population which is essential in bio-film formation \cite{chai2008bistability}, sporulation \cite{siebring2014repeated}, colony diversification \cite{koch2014evolution}, additional motility development \cite{bubendorfer2014secondary}, genomic island transfer \cite{ramsay2013widely}, quorum sensing \cite{perez2010heterogeneous,anetzberger2009heterogeneity} and many other context. In thorough time evolution experiments, the distribution of large scale inhomogeneities and co-existence of two steady states with drastically different concentration of proteins defined with sharp boundaries in a spatial distribution can be seen. Our model shows that transient spatial pattern in a diffusible cellular environment is observable in both 1D and 2D cellular array. Scanning the parameter space for different parameter values, we observed similar dynamics throughout the bistable region of the phase space. The effect of diffusion coefficient on the steadiness of the pattern of this motif is also established in our study. We quantified the time evolution in terms of island size growth.\\ Most of the studies related to pattern formation for biosystems are concerned with steady state patterns, and no significant investigation in transient dynamics is done so far. In some cases, the investigation of steady state pattern formation is valid for its biological relevance. For example, in \textit{Drosophila melanogaster}, the segment polarity network amplifies and maintains a periodic input, as a result an intrinsically stable pattern in output is formed. Though, in several examples, the biological pattern formation could be highly dynamic. Transient spatial pattern, which can be seen in the intermediate stages are often ignored as they do not represents the \lq final\rq  or \lq end\rq  state of the system. But, in study of the dynamical processes going on in living systems, these steps are highly important. Even in case of developement, where steady state patterns are considered to be benchmark, Conrad Hal Waddington, mentioned in “The Strategy of the Genes”, 1957 \cite{waddington2014strategy}
\begin{quote}
    In the study of development we are interested not only in the final state to which the system arrives, but also in the course by which it gets there.
\end{quote}
Thus, phenotypes should be described as emergent result of a continuous transition process between the patterns along with the time  required for these transitions, rather than the study of the final state \cite{jaeger2012inheritance, fusco2001many}. Some recent work, however beautifully described the importance of transient pattern formation and its importance in developmental pathway and cell fate determination, including gap gene model analysis  \cite{surkova2009canalization},  dorso-ventral
patterning of the vertebrate neural tube \cite{dessaud2007interpretation, balaskas2012gene}, vertebrate somitogenesis in arthropods \cite{cooke1976clock}, toggle switch \cite{verd2014classification} and many more. Increase in transient time with the increase in system size \cite{crutchfield1988attractors}, dependency of the temperature of the system, the role of evaporation in pattern formation, more specifically, in determining complexity of the spatial patterns are among some significant mathematical studies \cite{hamann2012self}.
\\Despite of these studies on transient pattern formation in biological systems, a few of them verifies the effect of growth on pattern formation \cite{crampin2001reaction, chaplain1999growth, lacalli1981dissipative}. 
Our work underscores the importance of considering host aspects while considering pattern formation, which emerges as a consequence of growth modulated feedback. This model particularly stands for those systems where expression of a protein is repressive to cell growth. While in every study related to mathematical modeling of synthetic circuits, one cell with a motif of interest is considered, the experimental conditions differ from that most of the time, which deals with a population of cell culture. In this condition, bistability of individual cells get affected by obvious diffusion from surrounding neighbor cells, and the dynamics becomes much more richer. The inherent stochasticity associated with reaction-diffusion gives rise to emergence of spatial fluxes, that is drastically different from the immediate homogenization. 
\\In the process of cell biology and bacterial phenotypes, the study of transient patterns helps in understanding the dynamical pathways for transitions. Despite of starting from a more or less homogeneous condition, while achieving phenotypic heterogeneity, the system gives rise and maintain large scale inhomogeneities and gradients. Here, it is important to mention that chemical gradients plays an important role in pattern formation and spatial phenotypic diversity helps an organism to determine corresponding pathway. To take this into account, in this work, we have considered different possible initial conditions as well, and analyzed the diversity of pattern formation. 
Though the patterns we observed for a range of diffusion coefficients, are transient, but the patches are also very persistent, even with random initial conditions. In case of multi-agent systems, the lifetime of the system could be comparable to the transition time \cite{hamann2012self}, making transient studies are immensely relevant there. 
It is important to note that these transient spatial structures, which are not stable from the mathematical point of view, might become stabilised by conjugate biological factors which are not being explicitly considered in mathematical calculations \cite{chatterjee2016eigenfunction} which opens up scope of experimental verification of our results.

\section*{Acknowledgement}
\noindent PC and SG acknowledge the support  by DST-INSPIRE, India, vide sanction Letter No. DST/INSPIRE/04/2017/002765 dated- 13.03.2019.

\section{Appendix }
\label{sec:sample:appendix}
\subsection{Diffusion in one dimension}
\begin{figure}
    \centering
    \includegraphics[width=\textwidth]{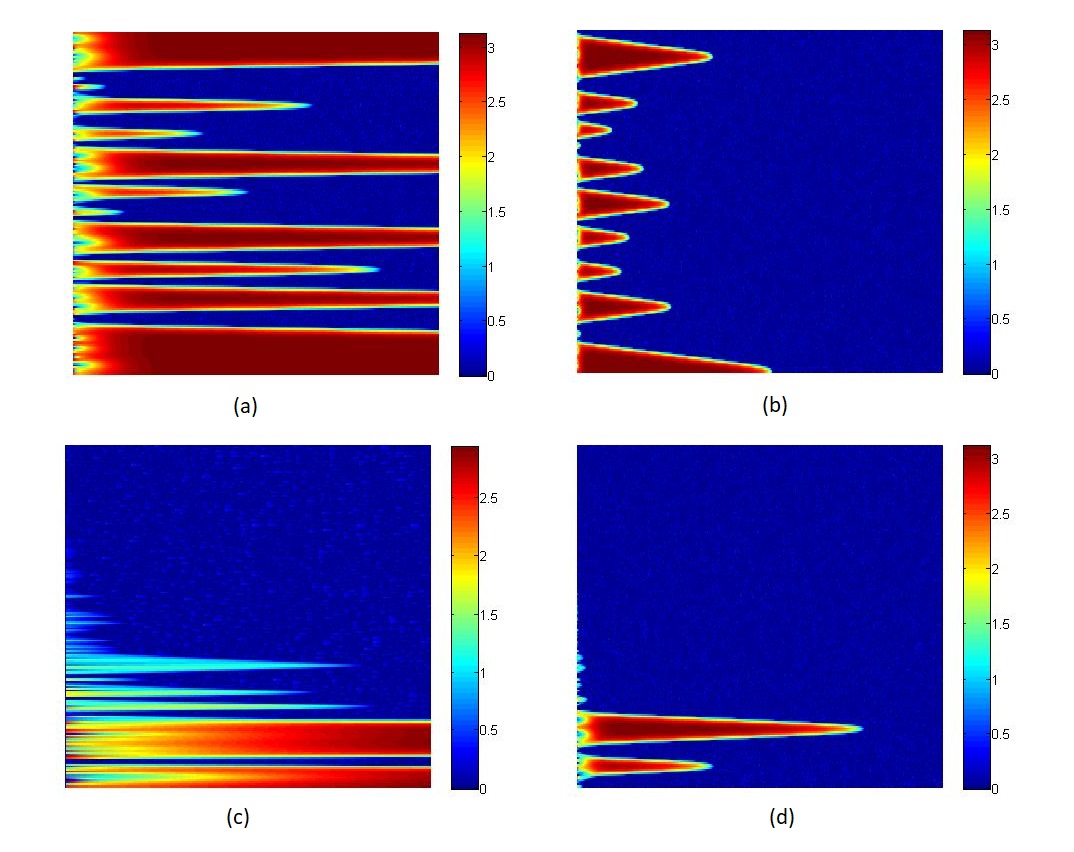}
    \caption{Diffusion in one dimension. Diffusion coefficient is $1$. Other parameter values are  $\alpha= 8.5, \gamma= 1.2, \phi=4, \beta= 0.01, \delta=0.03, \Delta=1$. $x$ axis represents time and $y$ axis represents positional information of the concentration of protein $U$. With exponential positive $x^2$ initial condition. Pattern after different time instants (a) 40 and (b) 300. With exponential negative $x^2$ initial condition, pattern  after different time instants (c) 5 and (d) 100.}
    \label{1dex}
\end{figure}

We chose some more initial conditions including positive and negative exponential function to test diffusion pattern in one dimension. The corresponding initial condition is given by Eq. \ref{appen} and diffusion patterns are given in Fig. \ref{1dex}.
\begin{equation}\label{appen}
    U_{initial}=k_1\: exp(\pm k_2\:x^2)
\end{equation}
$k_1$ is a number randomly chosen between the range of $0$ to $3$. $k_2$ is $0.0001$ for negative exponential condition and $0.000001$ for positive exponential. 
\subsection{Diffusion in two dimension}
\begin{figure}
    \centering
    \includegraphics[width=\textwidth]{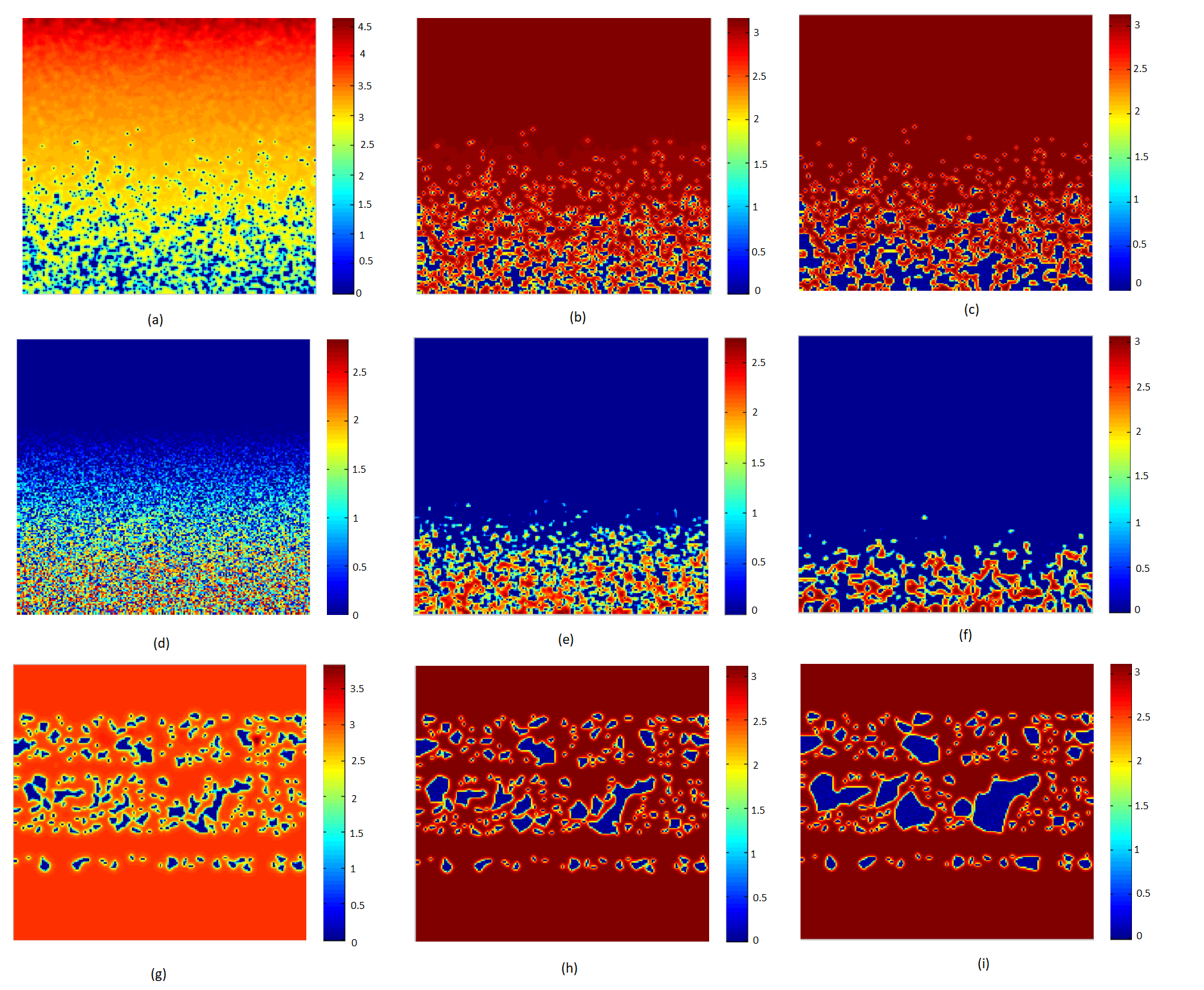}
    \caption{Diffusion in two dimension. Different parameter values are $\alpha= 8.5, \gamma= 1.2, \phi=4, \beta= 0.01, \delta=0.03, \Delta_U=1$. With exponential positive $x^2$ function. Diffusion coefficient is $0.2$. Pattern after different time instants (a) 5, (b) 40, and (c) 100. With exponential negative $x^2$ function. Diffusion coefficient is $0.3$. Pattern after different time instants (d) $0.25$, (e) 3, and (f) 8. With tangential initial condition.  Diffusion coefficient is $0.3$. Pattern after different time instants time (g) 50, (h) 85, and (i) 200.}
    \label{2dtan}
\end{figure}

We further have tested the condition mentioned in Eq. \ref{appen} and a tangential functions to test the diffusion pattern in two dimension. Here, for exponential functions 
$k_1$ is a number randomly chosen between the range of $0$ to $3$ and $k_2$ is $0.0001$. 
And the tangential initial condition is given by the equation below.
\begin{equation*}
    U_{initial}= k_1\: tan\frac{\pi\:x}{k_2}
\end{equation*}
 Here, value of $k_1$ is chosen randomly between $0$ to $3$, $k_2$ is $10$. The diffusion pattern can be seen in Fig. \ref{2dtan}.
\section{Supplementary Videos}
\begin{itemize}
    \item \href{https://youtu.be/uDBGO_LktsM}{Two dimensional random initialization:} 
    Video of the simulation for random initial condition, in presence of host-circuit interaction as shown in Fig. \ref{hstckt} Panel II. 
    \item \href{https://youtu.be/seRna3lGzl8}{Two dimensional stochastic periodic initialization:}
    Video of the simulation for stochastic periodic initial condition in two dimension, for Fig. \ref{2dransin}.
\end{itemize}

\newpage
 \bibliographystyle{elsarticle-num} 
 \bibliography{cas-refs}





\end{document}